\documentclass[prb,twocolumn]{revtex4}
\usepackage{epsfig}
\usepackage{bm}

\begin{document}

\title{The Rashba Effect within the Coherent Scattering Formalism}
\author{G. Feve}
\altaffiliation{also at ENS Cachan, 61 avenue du president Wilson,
94235 Cachan Cedex, France}
\author{W. D. Oliver}
\email{woliver@stanford.edu}
\author{M. Aranzana}
\altaffiliation{also at ENS Cachan, 61 avenue du president Wilson,
94235 Cachan Cedex, France}
\author{Y. Yamamoto}
\altaffiliation{also at NTT Basic Research Laboratories, 3-1
Morinosato-Wakamiya Atsugi, Kanagawa, 243-01 Japan}
\affiliation{Quantum Entanglement Project, ICORP, JST \\ E. L.
Ginzton Laboratory, \\ Stanford University, Stanford, CA 94305}
\date{\today}

\begin{abstract}
The influence of spin-orbit coupling in two-dimensional systems is
investigated within the framework of the Landauer-B\"{u}ttiker
coherent scattering formalism. This formalism usually deals with
spin-independent stationary states and, therefore, it does not
include a spin-orbit contribution to the current. In this article,
we will rederive the coherent scattering formalism, accounting for
the changes brought about by the spin-orbit coupling. After a
short review of the features of spin-orbit coupling in
two-dimensional electron gases, we define the creation/annihilaton
operators in the stationary states of the spin-orbit coupling
Hamiltonian and use them to calculate the current operator within
the Landauer-B\"{u}ttiker formalism. The current is expressed as
it is in the standard spin-independent case, but with the spin
label replaced by a new label which we call the spin-orbit
coupling label. The spin-orbit coupling effects can then be
represented in a scattering matrix which relates the spin-orbit
coupling stationary states in different leads. This scattering
matrix is calculated in the case of a four-port beamsplitter, and
it is shown to mix states with different spin-orbit coupling
labels in a manner that depends on the angle between the leads.
The formalism is then used to calculate the effect of spin-orbit
coupling on the current and noise in two examples of electron
collision.
\end{abstract}

\maketitle

\section{Introduction}

Coherent electron transport through nanostructures in cryogenic
two-dimensional electron gas systems is inherently a quantum
mechanical phenomenon. Several experiments have demonstrated
certain aspects of this quantum behavior, whether it relies
directly on the wave nature of the electron and can be probed
through a current or conductance measurement (\textit{e.g.},
quantized resistance in the quantum Hall effect \cite{QHE},
conduction modes of a quantum point contact \cite{QPC1,QPC2}), or
on the particle nature and quantum statistics of the electrons and
can be probed through a noise measurement (\textit{e.g.}, Hanbury
Brown and Twiss-type experiment \cite{HBTH,HBTW}, electron
collision \cite{EC}, observation of the fractional charge in the
fractional quantum Hall effect \cite{FQHE1,FQHE2}). In addition,
2D electron gases could be used to study the fundamental non-local
features of quantum mechanics through electron entanglement
\cite{EE,XA,NATO,BI,LOSS,WILL}. All these experiments can be
successfully explained within the coherent scattering formalism
\cite{LB,LA}, a theoretical tool describing coherent and
non-interacting electron transport. This formalism relies on
spin-independent stationary states in the leads of the device and,
therefore, describes spin-independent transport. Although it is
possible to add a local spin-dependent effect directly in the
scattering matrix, it is not possible in general to take into
account spin effects occurring over the whole system. One
potentially important spin effect occurring in the leads of the
conductor is spin-orbit coupling. Any electric field in the
reference frame of the laboratory generates a magnetic field in
the moving electron reference frame, coupling the electron's
orbital degrees of freedom with its spin. One can find several
sources of electric fields in semiconductors. In three dimensional
crystals, the periodic crystal potential generates the Dresselhaus
effect \cite{DRESS}, which induces a spin-splitting of the
conduction band that is proportional to $k^{3}$. In
two-dimensional systems, the dominant term results from the
asymmetry of the in-plane confining potential. First introduced by
Rashba \cite{RE2}, this effect causes a spin-splitting \cite{RE}
proportional to $k$, and it depends on the strength of the applied
electric field. Recently, there has been a growing interest in
electronic devices which rely on the spin properties of the
electrons. These spin-dependent devices may be influenced by
spin-orbit coupling effects, or may even rely on it as, for
example, in a coherent version of a spin-polarized field effect
transistor \cite{DA,MIR}. Therefore, it may be useful to include
spin-orbit coupling in the coherent scattering formalism.

\section{Spin-orbit coupling in two-dimensional electron gases}
\begin{figure}[h]
\epsfig{file=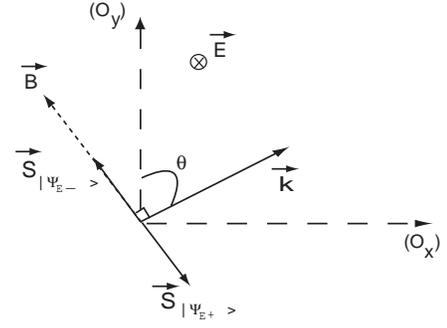,height=1.7in} \caption{Direction of the
virtual magnetic field and spin. An electron with wave vector
$\mathbf{k}$ in the presence of an electric field $\mathbf{E} = -
E_{0} \;\mathbf{u_{z}}$ feels a virtual field
$\mathbf{B}(\mathbf{k}, \mathbf{E})$ perpendicular to
$\mathbf{k}$.} \label{Fig1}
\end{figure}

\begin{figure}[h] \epsfig{file=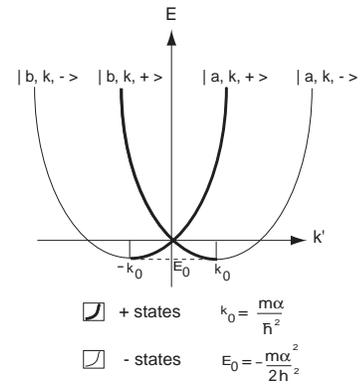,height=2in}
\caption{Energy dispersion diagram with SO coupling. The
$-\eta_{\sigma} k_{0} \leq k' < \infty$ range (a states) is
deduced from Eq.~(\ref{eq3}) with $k'=k$. The $-\infty < k' <
\eta_{\sigma} k_{0}$ range (b states) is then deduced from mirror
symmetry ($k'=-k$).} \label{Fig2}
\end{figure}

\begin{figure}[h]
\epsfig{file=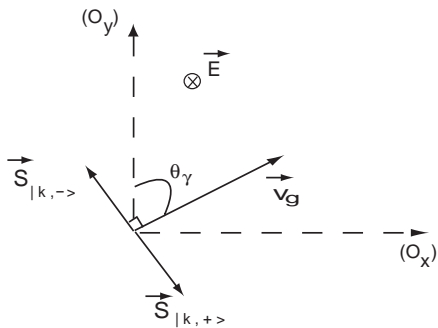,height=1.7in} \caption{Direction of
propagation and direction of the spin} \label{Fig3}
\end{figure}

The spin-orbit coupling is governed by the following spin-orbit
Hamiltonian \cite{QM}, which is obtained using an expansion in
$v/c$ of the Dirac equation
\begin{equation}
\hat{H}_{SO} = \frac{\hbar}{(2 m_{0} c)^{2}} \bm{\nabla} V  (
\bm{\hat{\sigma}} \times \mathbf{\hat{p}}) ,
\end{equation}
where $m_{0}$ is the free electron mass, $\mathbf{\hat{p}}$ is the
momentum operator,
$\bm{\hat{\sigma}}=(\hat{\sigma_{x}},\hat{\sigma_{y}},\hat{\sigma_{z}})
$ are the Pauli spin matrices , $V$ is the electrostatic
potential, and $\bm{\nabla}$ is the gradient operator so that
$-\bm{\nabla} V$ is the electric field. Electron transport in the
presence of an electric field results in a spin-orbit effect which
couples the electron spin and orbital degrees of freedom through
the $\bm{\hat{\sigma}} \times \mathbf{\hat{p}}$ term. Here, we
will neglect the bulk effect arising from the periodic crystal
potential (Dresselhaus effect) and consider only the effect caused
by the asymmetry of the confining quantum well (Rashba effect), as
it is stronger than the Dresselhaus effect in most two-dimensional
heterostructures \cite{GASB,ALSB,DA2}. Let z be the direction of
confinement, perpendicular to the plane of motion. The asymmetry
of the confining potential along the z-direction results in a
non-zero electric field along the z-axis ($\mathbf{E}=
-E_{0}\;\mathbf{u_{z}}$ throughout the article, where $u_{z}$
represents a unit vector in the z-direction). The spin-orbit
coupling term of the Hamiltonian can be written
\begin{equation}
\hat{H}_{SO} = \frac{\alpha}{\hbar}( \hat{\sigma} \times
\hat{p})_{z} = i \alpha (\hat{\sigma}_{y} \frac{
\partial}{\partial_{x}} - \hat{\sigma}_{x} \frac{\partial}{\partial
_{y}}) ,
\end{equation}
where $\alpha$ is the spin-orbit coupling constant and depends on
the strength of the electric field. It takes values in the range
$1$ to $10 \times 10^{-10}$ eV cm for a large variety of systems
(InAs/GaSb \cite{GASB}, InAs/AlSb \cite{HEIDA},
In$_{x}$Ga$_{1-x}$As/In$_{x}$Al$_{1-x}$As \cite{DA2,NITTA} and
GaAs/Al$_{x}$Ga$_{1-x}$As \cite{GAAS}) depending on the shape of
the confining well. Using the standard effective mass
approximation, we can deduce the system Hamiltonian as the free
Hamiltonian plus the spin-orbit coupling Hamiltonian
\begin{equation}
\hat{H} = \frac{\hat{p_{x}}^{2} + \hat{p_{y}}^{2}}{2 m} -
\frac{\alpha}{\hbar}(\hat{\sigma_{y}} \hat{p_{x}} -
\hat{\sigma_{x}} \hat{p_{y}}) ,
\end{equation}
where m is always taken to be the effective mass. Since the
operators $\hat{p_{x}}$ and $\hat{p_{y}}$ commute with $\hat{H}$,
we can search for eigenstates of the form
\begin{equation}
\left| \psi \right> = e^{ i( k_{x}x + k_{y}y)} \;[ \beta \left|
\uparrow\right> + \gamma \left| \downarrow \right>] ,
\end{equation}
where $\left| \uparrow\right>$ and $ \left| \downarrow\right> $
label the up and down states of the z-component of the spin. We
can now diagonalize the Hamiltonian
\begin{eqnarray}
\hat{H} & = & \left( \begin{array}{cc}
             \frac{\hbar^{2} k^{2}}{2m}      &  \alpha k_{y} + i \alpha k_{x} \\
             \alpha k_{y} - i \alpha k_{x}   &  \frac{\hbar^{2} k^{2}}{2m}  \\
           \end{array} \right)
\end{eqnarray}
in this spin subspace. The eigenvalues are $E(k) = \frac{\hbar
k^{2}}{2 m} \pm \alpha k $, and the associated eigenfunctions are
\begin{eqnarray}
 \left| \Psi _{E_{+}} \right> & = &
\frac{e^{i (k_{x}x + k_{y}y)}}{\sqrt{2}} [ e^{i \frac{\theta}{2}}
\left| \uparrow \right> + \;e^{-i \frac{\theta}{2}} \left|
\downarrow \right> ] \label{eq1} \\
 \left| \Psi _{E_{-}} \right> & = &
\frac{e^{i (k_{x}x + k_{y}y)}}{\sqrt{2}} [e^{i \frac{\theta}{2}}
\left| \uparrow \right> - \;e^{-i \frac{\theta}{2}} \left|
\downarrow \right> ] \label{eq2}
\end{eqnarray}
where $\theta$ is the angle between $\mathbf{k}=(k_{x},k_{y})$ and
the y-axis (see Fig.~\ref{Fig1}). The electrons feel a virtual
magnetic field in the 2D plane in a direction perpendicular to
$\mathbf{k}$.
The spin is aligned or antialigned to this field (see
Fig.~\ref{Fig1}) , so that $\theta (\mathbf{k}, \mathbf{S}_{\left|
\Psi_{E_{-}} \right>}) = \pi/2$, and $\theta (\mathbf{k},
\mathbf{S}_{\left| \Psi_{ E_{+}} \right>}) = -\pi/2$, where
$\theta (\mathbf{k}, \mathbf{S}_{\left| E_{-} \right>})$ is the
angle between $\mathbf{k}$ and the spin $\mathbf{S}$ in the state
$\left| \Psi_{E_{-}}\right>$. The amplitude of the magnetic field
depends on the velocity of the electron and vanishes for $k=0$,
preventing a possible spin polarization in the system. The
spin-orbit splitting is usually small compared to the kinetic
energy of the electrons ( $0.15$ meV $\leq \alpha k_{f} \leq 1.5$
meV for $E_{f} =14$ meV). Following Ref.~\onlinecite{LB}, we will
introduce a transverse confinement in the leads of the conductor,
allowing us to address the longitudinal transport modes for each
given transverse mode. Focusing on one lead, $\gamma$, we will
make two simplifying assumptions. First, we consider only a single
independent transverse mode. Second, we neglect the 1D SO coupling
effect that this transverse confining potential could create,
since, to our knowledge, there is no experimental verification of
this effect, and it is estimated to be much smaller than the
Rashba effect \cite{MO}. Within these approximations, we can use
our previous analysis to deduce the eigenstates and the associated
energy dispersion diagram (see Fig.~\ref{Fig2}), with $k$ lying in
the direction $x_{\gamma}$ of the lead (making the angle
$\theta_{\gamma}$ with the y axis). We will now introduce three
labels for the eigenstates that will prove useful in writing the
creation and annihilation operators of these states. $\epsilon = a
$ or $b$ labels the direction of propagation from the sign of the
group velocity $v_{g}$ ($a$ if $sgn(v_{g}) >0 $, b otherwise). The
parameter $k$ labels the longitudinal mode wavevector. The SO
coupling label $\sigma \equiv \pm$ designates the two different
branches of the energy dispersion diagram for a given k ($+$
branch and $-$ branch in Fig.~\ref{Fig2}).
Using these labels, we find the following eigenstates and
eigenvalues of the system from Eq.~(\ref{eq1}) and (\ref{eq2})
\begin{eqnarray}
\left| \epsilon, k, \sigma \right> & = & \Phi(y_{\gamma})
\frac{e^{i \eta_{\epsilon} k x_{\gamma}}}{\sqrt{2}} [ e^{i
\frac{\theta_{\gamma}}{2}} \left| \uparrow\right> +
\;\eta_{\epsilon} \eta_{\sigma} e^{-i \frac{\theta_{\gamma}}{2}}
\left| \downarrow\right> ] \label{eq3} \nonumber \\* E & = &
\frac{\hbar k^{2}}{2m} + \eta_{\sigma} \alpha k \;\;\;\;\;
-\eta_{\sigma} k_{0} \leq k < \infty
\end{eqnarray}
where $\eta_{\epsilon} = +1 (-1)$ when $\epsilon = a (b)$, and
$\eta_{\sigma} = +1 (-1)$ when $\sigma = + (-)$.
$\Phi(y_{\gamma})$ is the normalized transverse wavefunction for
the transverse mode under consideration. By convention, $k$ is
only taken in the range $-\eta_{\sigma} k_{0} \leq k < \infty $
for both the $a$ and $b$ states. We use $\eta_{\epsilon} = \pm 1$
to parameterize explicitly the appropriate wavevector range for
the $\epsilon = a (b)$ propagation direction in the eigenstates'
phase factor $e^{i \eta_{\epsilon} k x}$. The eigenvalues in
Fig.~\ref{Fig2} can be deduced by mirror symmetry. This convention
allows us to track the propagation direction throughout the
calculation. We also notice in Fig.~\ref{Fig2} that, for a given
energy, the corresponding $+$ and $-$ states with same label
$\epsilon$ do not have the same wavevector:
\begin{equation}
k(E,-) - k(E,+) = \Delta k = 2 k_{0} = \frac{2 m
\alpha}{\hbar^{2}} .\label{eq4}
\end{equation}
The states with $\sigma= \pm$ have their spin perpendicular to the
direction of propagation $\mathbf{v_{g}}$ and in opposite
directions, so that $\theta(\mathbf{v_{g}}, \mathbf{S}_{\left| k,
-\right>}) = \pi/2$ and $\theta( \mathbf{v_{g}},
\mathbf{S}_{\left| k, +\right>}) = -\pi/2$ (see Fig.~\ref{Fig3}).
It is important to notice that the $\sigma=+ (-)$ states do not
coincide with the $\left| \Psi_{E_{+}}\right> \left( \left|
\Psi_{E_{-}}\right> \right)$ states, as it is $\mathbf{v_{g}}$,
and not $\mathbf{k}$, which determines the direction of the spin
(\textit{e.g.}, in Fig.~\ref{Fig3}, the spin of the $+$ state is
perpendicular to and on the right side of $\mathbf{v_{g}}$). For
example, in Fig.~\ref{Fig2}, the two definitions are consistent
for states with positive (negative) $k'$ and group velocity
$v_{g}$. However, they are inconsistent for positive (negative)
$k'$ and negative (positive) $v_{g}$. We choose here to link the
direction of the spin to $\mathbf{v_{g}}$ (rather than
$\mathbf{k'}$) through the label $\sigma=\pm$ (rather than
$E_{\pm}$), because it is the group velocity which determines the
direction of the current. Note that in the case without SO
coupling, no differentiation between $v_{g}$ and $k'$ is necessary
since they always share the same sign.

\section{Spin-Orbit coupling and coherent scattering formalism}
The Landauer-B\"{u}ttiker formalism relies on a second
quantization formulation of quantum mechanics. The current is
expressed as a function of the field operator, which is expanded
in the basis of the scattering states using the operators that
create and destroy electrons in the leads of the conductor. A
scattering state is a coherent sum of an incident wave in one lead
and the outgoing waves it generates in all the other leads. The
amplitude of each outgoing wave is given by the scattering matrix
$S$, whose elements depend on the properties of the scatterer. As
our final goal is to study the fluctuations of the current, it is
important to find a simple way to express the time dependence of
the current operator. This leads us to choose the stationary
states in each lead for the purpose of expanding the field
operator. In our rederivation of the coherent scattering
formalism, we will follow the same analysis as in
Ref.~\onlinecite{LB}, but introducing two important changes: we
will rederive the expression of the current operator as a function
of the field operator, adding a new contribution coming from
spin-orbit coupling, and we will expand the field operator in the
new stationary basis of the SO coupling Hamiltonian.

\subsection{Calculation of the current operator}
One way to find the current operator for non-interacting electrons
is to use the single-particle Schr\"{o}dinger equation
\begin{equation}
i \hbar \frac{\partial \Psi}{\partial t}  =  -\frac{\hbar ^{2}}{ 2
m} \;( \;\frac{\partial \Psi}{\partial x ^{2}} + \frac{\partial
\Psi}{\partial y^{2}} \;)   + i \alpha \;(\; \sigma_{y}
\frac{\partial \Psi}{\partial x} - \sigma _{x} \frac{\partial
\Psi}{\partial y} \; ) ,
   \label{eq5}
   \end{equation}
where $\Psi$ is a two-components spinor. The current operator
$\mathbf{\hat{j}}$ can be extracted from the conservation of
charge equation $\frac{\partial}{\partial t} [ e \Psi^{\dag} \Psi]
+ \bm{\nabla}.\mathbf{j} = 0$; using Eq.~(\ref{eq5}) and its
adjoint,
\begin{eqnarray}
j_{x}& = & \frac{e \hbar}{2 m i} \Big[\; \Psi^{\dag}
\frac{\partial \Psi}{\partial x} - \frac{\partial \Psi
^{\dag}}{\partial x} \Psi \;\Big] - \frac{e \alpha}{\hbar}\;
\Psi^{\dag} \sigma_{y} \Psi \\ j_{y} &=& \frac{e \hbar}{2 m i}
\Big[\; \Psi^{\dag} \frac{\partial \Psi}{\partial y} -
\frac{\partial \Psi ^{\dag}}{\partial y} \Psi \;\Big] + \frac{e
\alpha}{\hbar} \; \Psi^{\dag} \sigma_{x} \Psi
\end{eqnarray}
We can identify the usual kinetic term of the current density
\begin{equation}
\mathbf{j_{K}}= \frac{e \hbar}{2 m i} \Big[\; \Psi^{\dag} \nabla
\Psi - \nabla \Psi ^{\dag} \Psi\;\Big] . \label{eq6}
\end{equation}
Note that SO coupling adds a new contribution proportional to
$\alpha$ that we call the SO coupling current density
$\mathbf{j_{SO}}$,
\begin{equation} \mathbf{j_{SO}} = - \frac{e \alpha}{\hbar} \;
\Psi^{\dag} \sigma_{y} \Psi \;\mathbf{u_{x}} + \frac{e
\alpha}{\hbar} \; \Psi^{\dag} \sigma_{x} \Psi \;\mathbf{u_{y}} .
\label{eq7}
\end{equation}
In the framework of second quantization, $\Psi$ becomes a field
operator, and we have to expand it using a convenient basis: the
stationary states.

\subsection{Expansion of the field operator}
\begin{figure}[h] \epsfig{file=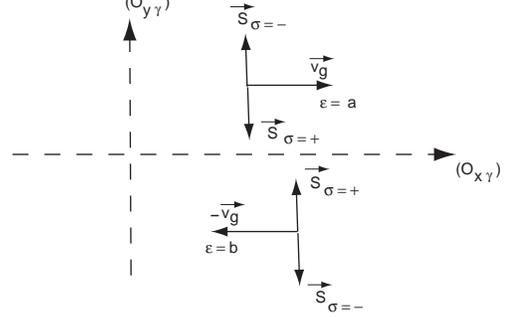,height=1.7in}
\caption{Direction of the spin of $\chi^{\epsilon}_{\sigma}$}
\label{Fig4}
\end{figure}

The first step is to define the creation (annihilation) operators
for the SO coupling stationary states. We first introduce the
creation operators in the old spin basis: $\hat{a}^{\dag}_{\gamma
k \uparrow}$ ($\hat{b}^{\dag}_{\gamma k \downarrow}$) creates an
incoming (outgoing) electron with spin up (down) in the first mode
of lead $\gamma$ with momentum $k$, in the state $\left| a, k,
\uparrow \right>_{\gamma}$ ($\left| b, k, \downarrow
\right>_{\gamma}$), where $\left| \epsilon, k, \uparrow
\right>_{\gamma} = e^{i \eta_{\epsilon} k x_{\gamma}} \left|
\uparrow \right>$. These standard operators satisfy the
anticommutation relation $\Big[\hat{a}_{\gamma, k,
s}\;,\;\hat{a}^{\dag}_{ \beta ,k', s'}]_{+} = \delta_{k k'}
\delta_{\gamma \beta} \delta_{s s'}$, where $s$ labels the spin of
the electron. Through Eq.~(\ref{eq3}) we introduce
$\hat{a}^{\dag}_{\gamma k \sigma}$, which creates an incoming
electron in lead $\gamma$ with momentum $k$ and spin-orbit
coupling label $\sigma$ (not to be confused with the spin $s$)
satisfying the relation
\begin{equation}
\hat{a}^{\dag}_{\gamma k \sigma} = \frac{1}{\sqrt{2}} [ e^{i
\frac{\theta_{\gamma}}{2}} \hat{a}^{\dag}_{\gamma k \uparrow} +
\eta_{\sigma} e^{-i \frac{\theta_{\gamma}}{2}}
\hat{a}^{\dag}_{\gamma k \downarrow}] .
\end{equation}
Knowing the anticommutation relationships in the spin basis, we
can calculate it in our new stationary basis to be
\begin{equation}
\Big[ \hat{a}_{k, \gamma, \sigma } \; , \; \hat{a}^{\dag}_{k',
\beta, \sigma'} \Big]_{+}  =  \delta_{k k'} \delta_{\alpha \beta}
\delta_{\sigma \sigma'} ,
\end{equation}
where we used the relation $ \eta_{\sigma}\eta_{\sigma'} =  2
\delta_{\sigma \sigma'} - 1$. This result justifies our labeling
of the SO coupling eigenstates and the use of the spin-orbit
coupling label $\sigma$ instead of the spin. The stationary states
form a complete basis, and we can use them to expand the field
operator in lead $\gamma$. From now on, we will choose
$x_{\gamma}$ and $y_{\gamma}$ as the reference axis (so that
$\theta_{\gamma} = \frac{\pi}{2}$), specifically tracking all of
the angular dependence in the scattering matrix.
\begin{equation}
\hat{\Psi}(x,y)_{\gamma}  =  \sum_{\sigma , \epsilon} \sum_{k = -
\eta_{\sigma}\;k_{o} }^{\infty}  \hat{\epsilon}_{\gamma, k,
\sigma} \;\chi^{\epsilon}_{\sigma}   \; \frac{e^{i \eta_{\epsilon}
kx} \Phi (y)}{\sqrt{L}}
\end{equation}
where $\hat{\epsilon} = \hat{a} (\hat{b})$ is the annihilation
operator in the incoming(outgoing) states and
$\chi^{\epsilon}_{\sigma}$ is a two-component spinor obtained from
Eq.~(\ref{eq3}), setting $\theta_{\gamma} = \frac{\pi}{2}$
\begin{equation} \left| \epsilon, k , \sigma \right>_{\gamma} =
\frac{e^{i \eta_{\epsilon} k x}\;\Phi(y)}{\sqrt{L}} \underbrace{
\left[
\begin{array}{c} \frac{1+ i }{2} \\ \eta_{\epsilon}
\eta_{\sigma} \frac{1-i }{2}  \\
\end{array} \right]}_{ \chi^{\epsilon}_{\sigma}} .
\label{eq8}
\end{equation}
As shown in Fig.~\ref{Fig4}, its spin depends on the SO coupling
state ($\sigma$) and on the direction of propagation ($a$ or $b$).
At this point, it is more convenient to go from a discrete sum in
$k$-space to a continuous integral in energy by defining the
operators $ \hat{a}^{\dag}(E)$, which create electrons in the
energy continuum. As our labeling defines a one-to-one
correspondence between ($\epsilon, k, \sigma$) and ($\epsilon, E,
\sigma$), we can define unambiguously $\hat{a}^{\dag}_{\gamma
\sigma}(E) = \sqrt{\rho(E)} \hat{a}^{\dag}_{\gamma k \sigma}$, so
that
\begin{equation}
\Big[ \hat{a}_{ \gamma, \sigma }(E) \; , \; \hat{a}^{\dag}_{
\beta, \sigma'}(E') \Big]_{+} = \delta _{\gamma \beta}
\delta_{\sigma \sigma'} \delta (E-E')
\end{equation}
where $\rho(E)$ is the density of states at energy E . \textit{A
priori}, $\rho(E)$ could depend on the spin-orbit coupling label
$\sigma$, but one can check in Appendix~\ref{apa} that it is not
the case. Replacing the discrete sum over $k$ by an integral over
$E$ and introducing the new operators and their time-dependence
$\hat{a}(E,t) = e^{-i\frac{E}{\hbar} t} \hat{a}(E)$, we find
\begin{eqnarray}
\hat{\Psi}(x,y,t)_{\gamma} = \frac{1}{\sqrt{2 \pi}} \sum_{\sigma ,
\epsilon } \int_{E_{0}}^{\infty} \frac{dE}{\sqrt{\hbar
\;v_{g}(E)}} \hat{\epsilon}_{\gamma, \sigma}(E)
\;\chi^{\epsilon}_{\sigma}\Phi(y) \nonumber \\ \times \;e^{i
\eta_{\epsilon} k(E,\sigma) x} e^{-i \frac{E}{\hbar}t} \qquad
\end{eqnarray}
with $E_{0} \equiv -\frac{m \alpha^{2}}{2 \hbar^{2}}$, and $k$ has
been replaced by $k(E,\sigma)$ to remind us that, for a given
energy, $k$ depends on the spin-orbit coupling label $\sigma$.

\subsection{Current in the spin-orbit coupling basis}
The current in lead $\gamma$ is given by
\begin{equation}
\hat{I}_{\alpha} = \int dy \hat{j}_{x} = \int dy (\hat{j}_{x K} +
\hat{j}_{x SO} ) \equiv \hat{I}_{K} + \hat{I}_{SO} .
\end{equation}
We will begin by calculating the usual kinetic term $\hat{I}_{K}$.
Using Eq.~(\ref{eq6}), we have
\begin{widetext}
\begin{equation}
\hat{I}_{K}(t) =  \sum_{\sigma \sigma'} \sum_{\epsilon \;
\epsilon'} \int dE dE' \;\hat{I}_{K\;\sigma \sigma'}^{\;\epsilon
\; \epsilon'}(E, E') \;\hat{\epsilon}^{+ }_{\gamma \sigma}(E)
\;\hat{\epsilon'}_{\gamma \sigma'}(E') \;e^{i \frac{E -
E'}{\hbar}t} .\label{eq9}
\end{equation}
\begin{equation}
\hat{I}_{K\;\sigma \sigma'}^{\;\epsilon \; \epsilon'}(E, E')
\;=\;\frac{e \hbar}{2 m i}\; \frac{i \chi_{\sigma}^{\epsilon
^{\dag}} \chi_{\sigma'}^{\epsilon'}}{h \sqrt{v_{g}(E)
v_{g}(E')}}[\;\eta_{\epsilon} k(E,\sigma) + \eta_{\epsilon'}
k(E',\sigma')\;]\;e^{i\;(\eta_{\epsilon'} k(E',\sigma') -
\eta_{\epsilon} k(E,\sigma ))x} .
\end{equation}
\end{widetext}
We notice from Eq.~(\ref{eq9}) that the $\omega $
frequency of the current is given by $E=E' + \hbar \omega$.
Following Ref.~\onlinecite{LB}, we calculate the current and noise
in the zero-frequency limit, and make the approximation
$k(E',\sigma) \approx k(E,\sigma)$ (as $E \approx E'$) for
electrons having the same SO coupling label, and, using
Eq.~(\ref{eq4}), $k(E',\sigma') = k(E,\sigma) + \eta_{\sigma}
\Delta k$ for electrons having different SO coupling labels. From
Eq.~(\ref{eq8}), we also have, (with $\delta_{\epsilon \neq
\epsilon'} = 1$ if $\epsilon \neq \epsilon', 0$ otherwise)
\begin{equation}
\chi_{\sigma}^{\epsilon ^{\dag}} \chi_{\sigma'}^{\epsilon'} =
\delta_{\epsilon \epsilon'}\delta_{\sigma \sigma'} +
\delta_{\epsilon \neq \epsilon'}\delta_{\sigma \neq \sigma'}.
\end{equation}
This result is consistent with the fact that only electrons with
the same direction of propagation and the same spin-orbit coupling
label, or opposite direction of propagation and opposite
spin-orbit coupling label, have the same spin.

\begin{widetext}
\qquad \\
\indent In the $\sigma =\sigma'$ case, implying $\epsilon =
\epsilon'$ and $k(E',\sigma') \approx k(E,\sigma)$, we find
\begin{equation}
\hat{I}_{K\;\sigma \sigma'}^{\;\epsilon \; \epsilon'}(E,
E')\mid_{\sigma = \sigma'} \;=\; \eta_{\epsilon} \frac{e}{h} \;
\frac{k(E,\sigma)}{k(E,+) + \frac{m \alpha}{\hbar^{2}} } .
\label{eq10}
\end{equation}

In the $\sigma \neq \sigma'$ case, implying $\epsilon \neq
\epsilon'$ and $k(E',\sigma') \approx k(E,\sigma) + \eta_{\sigma}
\frac{2m\alpha}{\hbar^{2}}$, so that $\eta_{\epsilon} k(E,\sigma)
+ \eta_{\epsilon'} k'(E,\sigma') = -\eta_{\epsilon} \;
\eta_{\sigma} \frac{2 m \alpha}{\hbar^{2}}$, we find
\begin{equation}
\hat{I}_{K\;\sigma \sigma'}^{\;\epsilon \; \epsilon'}(E,
E')\mid_{\sigma \neq \sigma'} \;=\;- \eta_{\epsilon} \frac{e}{h}\;
\frac{\eta_{\sigma}\frac{m \alpha}{\hbar^{2}}}{k(E,+) + \frac{m
\alpha}{\hbar^{2}}}\; e^{-i \eta_{\epsilon} ( 2 k(E,\sigma) +
\eta_{\sigma} \Delta k )x} . \label{eq11}
\end{equation} If we only
consider the kinetic term of the current, a non-vanishing
contribution for $\epsilon \neq \epsilon'$ leads to terms like
$\hat{a}^{\dag}\hat{b}$ and $\hat{b}^{\dag}\hat{a}$ in the
expression of the current, corresponding to electrons propagating
in opposite directions.

Now, let us study the contribution of the spin-orbit coupling
current $\hat{I}_{SO}$. Using Eq.~(\ref{eq7}) to calculate
$\hat{I}^{\;\epsilon \;\epsilon'}_{SO \;\sigma \sigma'}$ in
Eq.~(\ref{eq9}) with $\scriptstyle K \rightarrow SO$, we find
\begin{equation}
\hat{I}^{\;\epsilon \;\epsilon'}_{SO \;\sigma \sigma'}(E,E') =
-\frac{e}{h} \;\frac{\frac{\alpha}{\hbar}}{ \sqrt{v_{g}(E)
v_{g}(E')}}\;\chi_{\sigma}^{\epsilon ^{\dag}} \sigma_{y}
\chi_{\sigma'}^{\epsilon'} \;e^{i(\eta_{\epsilon'} k(E',\sigma') -
\eta_{\epsilon} k(E,\sigma))x} .
\end{equation}
Using
$\eta_{\epsilon'}\eta_{\sigma'}=\eta_{\epsilon}\eta_{\sigma}$, we
have from Eq.~(\ref{eq8}) $\sigma_{y} \chi_{\sigma'}^{\epsilon'} =
-\eta_{\epsilon'} \eta_{\sigma'} \chi_{\sigma'}^{\epsilon'} $ and
$\chi_{\sigma}^{\epsilon ^{\dag}}\sigma_{y}
\chi_{\sigma'}^{\epsilon'}  = -\eta_{\epsilon} \eta_{\sigma} \;(
\delta_{\epsilon \epsilon'} \;\delta_{\sigma
\sigma'}\;+\;\delta_{\epsilon \neq \epsilon'}\; \delta_{\sigma
\neq \sigma'})$, so that
\begin{equation}
\hat{I}^{\;\epsilon \;\epsilon'}_{SO \; \sigma \sigma'}(E,E') =
\eta_{\epsilon}  \frac{e}{h} \; \frac{ \eta_{\sigma}
\;\frac{m\alpha}{\hbar^{2}}}{k(E,+) + \frac{m \alpha}{\hbar^{2}} }
\; \Big[\;\delta_{\epsilon \epsilon'} \;\delta_{\sigma \sigma'} +
\;e^{-i \eta_{\epsilon} ( 2 k(E,\sigma) + \eta_{\sigma} \Delta k
)x}\; \delta_{\epsilon \neq \epsilon'} \; \delta_{\sigma \neq
\sigma'}\;\Big] . \label{eq12}
\end{equation}
\end{widetext}
We can now calculate the value of the total current from
Eq.~(\ref{eq10}), (\ref{eq11}) and (\ref{eq12})
\begin{eqnarray}
\hat{I}^{\;\epsilon \;\epsilon'}_{\; \sigma \sigma'}(E,E') & = &
\hat{I}^{\;\epsilon \;\epsilon'}_{K\; \sigma \sigma'}(E,E') +
\hat{I}^{\;\epsilon \;\epsilon'}_{SO\; \sigma \sigma'}(E,E')
\nonumber \\ & = & \eta_{\epsilon} \frac{e}{h}\;\delta_{\epsilon
\epsilon'} \;\delta_{\sigma \sigma'} .
\end{eqnarray}
Although the spin-orbit coupling mixes electrons having different
directions of propagation when we consider only the kinetic term
of the current, these contributions cancel when we add the
spin-orbit coupling current $I_{SO}$, and we find the standard
formula for the current operator,
\begin{eqnarray}
\hat{I} \; = \; \frac{e}{h} \sum_{\sigma} \int dE\; dE' \;[\;
\hat{a}^{\dag}_{\gamma \sigma}(E)\;\hat{a}_{\gamma \sigma}(E')
\qquad \nonumber \\*  - \hat{b}^{\dag}_{\gamma
\sigma}(E)\;\hat{b}_{\gamma \sigma}(E')\;]\; e^{i
\frac{E-E'}{\hbar}t} .
\end{eqnarray}
In this formula, the definition of the $\hat{a}$($\hat{b}$) states
as states with positive(negative) group velocity, and not
necessarily positive(negative) wavevector, is consistent with the
fact that they carry the current in opposite directions. More
importantly, the final expression of the current is similar to the
one found in Ref.~\onlinecite{LB}, but with the spin index
replaced by the spin-orbit coupling label $\sigma = \pm$. The spin
related to this new index depends on the direction of propagation,
that is, the angle of the lead $\theta_{\gamma}$. Therefore, we
expect a $\theta_{\gamma}$ dependence in the scattering matrix
relating the spin-orbit coupling states in leads with different
orientations. As an example, we investigate the scattering matrix
in the case of a four-port beamsplitter (two input leads, two
output leads) used in electron collisions. The scattering matrix
relates the outgoing states in the outputs ($\hat{b} $ states) to
the incident states at the input ($\hat{a}$ states)
\begin{equation}
\hat{b}_{\gamma \sigma} = \sum_{\beta \sigma'} S^{\gamma
\sigma}_{\beta \sigma'} \; \hat{a}_{\beta \sigma'} . \label{eq13}
\end{equation}
In the spin-independent transport case, as the beamsplitter does
not act on the spin degrees of freedom, the scattering matrix does
not mix different spins (the scattering matrix is diagonal in each
subspace of spin: spin-up and spin-down). In this case the
reflection and transmission coefficients do not depend on the
spin. However, when we include spin-orbit coupling, the situation
is more complicated, because the spin associated with the
spin-orbit coupling label $\sigma$ is different in each lead (the
leads having different orientations). The conservation of the spin
at the beamsplitter then implies that we have a mixing of the
spin-orbit states (off-diagonal elements in the scattering
matrix), and this mixing becomes more important when the angle
between the leads increases.
\begin{figure}[h]
\epsfig{file=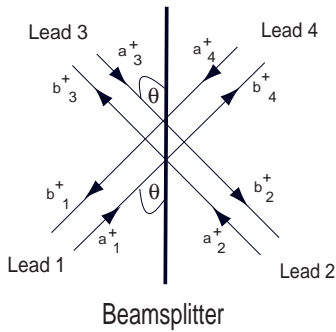,height=1.7in} \caption{Four-port
beamsplitter} \label{Fig5}
\end{figure}
A beamsplitter, with all leads oriented at the same angle $\theta$
(the case illustrated in Fig.~\ref{Fig5}) is investigated in
Appendix~\ref{apb}, and the following scattering matrix is found
after some simplifying assumptions
\begin{widetext}
 \samepage
 \begin{eqnarray}
 \left( \begin{array}{c}
    \hat{b}_{3+} \\
    \hat{b}_{3-} \\
    \hat{b}_{4+} \\
    \hat{b}_{4-}
    \end{array} \right)
= \underbrace{ \left( \begin{array}{cccc}
      r\cos\theta & i \; r \sin\theta & t & 0  \\
      i \; r \sin\theta & r\cos\theta & 0 & t  \\
      t & 0 & r\cos\theta & -i \;r \sin\theta  \\
      0 & t & - i \;r \sin\theta & r\cos\theta
      \end{array} \right) }_{S}
  \left( \begin{array}{c}
    \hat{a}_{1+} \\
    \hat{a}_{1-} \\
    \hat{a}_{2+} \\
    \hat{a}_{2-}
    \end{array} \right)
\label{eq14}
\end{eqnarray}
We note that the unitarity of the $S$ matrix requires $\left| r
\right| ^{2} + \left| t \right| ^{2} \equiv R + T = 1$ and $ r
t^{*} + tr^{*}
 =  0$. \\
\end{widetext}

\section{Spin-orbit coupling and noise in electron collisions}

We will now use the expression of the current operator and the
scattering matrix to calculate the current and noise in two
experiments using the four-port beamsplitter setup. The first is
an unpolarized electron collision, which has already been realized
experimentally (see Ref.~\onlinecite{EC}). The second is a
spin-polarized electron collision.

\subsection{Unpolarized electron collision}
This experiment is a collision at a 50/50 beamsplitter of
electrons from leads 1 and 2 which are biased at $E_{f} + eV.$ At
zero temperature and without spin-orbit coupling, each electron
with a specific spin in lead 1 has an identical partner in lead 2.
The Pauli exclusion principle then forbids us to find these two
electrons in the same output. We expect a total suppression of the
partition noise. With spin-orbit coupling, we do not expect this
result to be modified. It relies only on the full occupation of
the energy eigenstates, which is not modified at zero temperature
by SO coupling. We will confirm this by studying the collision of
four electrons at the same energy. The initial state is:
\begin{equation}
\left|\Psi_{i}\right>  =   \hat{a}^{\dag}_{k, +, 1}(E)\;
\hat{a}^{\dag}_{k', -, 1}(E)\; \hat{a}^{\dag}_{k, +,
2}(E)\;\hat{a}^{\dag}_{k', -, 2}(E)\;\left| 0\right> ,
\end{equation}
From now on, we will drop the $E$, $k$ and $k'$ labels for the
energy and the momentum of the electrons, associating a $+$($-$)
SO coupling state with $k$($k'$). The output state is found using
the new scattering matrix with SO coupling using Eq.~(\ref{eq13})
and Eq.~(\ref{eq14})
\begin{widetext}
\begin{eqnarray}
\left|\Psi_{f}\right> & = & \left[r\;[\;\cos\theta
\;\hat{b}^{\dag}_{+ 3} + i\sin\theta \; \hat{b}^{\dag}_{- 3} \;] +
t\;\hat{b}^{\dag}_{+ 4}\right]\;\left[r\; [\;i \sin\theta \;
\hat{b}^{\dag}_{+ 3} + \cos\theta \; \hat{b}^ {\dag}_{- 3}\;] +
t\;\hat{b}^{\dag}_{- 4}\right]\nonumber \\* &  &
\times\left[t\;\hat{b}^{\dag}_{+ 3} + r\;[\;\cos\theta\;
\hat{b}^{\dag}_{+ 4} - i\sin\theta \;\hat{b}^{\dag}_{- 4}
\;]\right]\; \left[t\;\hat{b}^{\dag}_{- 3} + r\;[\;- i \sin\theta
\; \hat{b}^{\dag}_{+ 4} + \cos\theta \;\hat{b}^{\dag}_{-
4}\;]\right] \left| 0\right>  \nonumber \\* &=& \hat{b}^{\dag}_{+
3}\hat{b}^{\dag}_{- 3}\hat{b}^{\dag}_{+ 4}\hat{b}^{\dag}_{- 4}
\left| 0\right>
\end{eqnarray}
 \end{widetext}
where we have used the relation $r^{4} + t^{4} -2 \;r^{2} \;t^{2}
= 1$. As expected, a full occupation of the two input states at
energy E leads to a full occupation of the two output states as
well. A calculation of the noise for the fully occupied outputs
would show a complete suppression of the partition noise. The only
difference with the spin-independent transport case is that the
colliding electrons at the same energy do not have the same
momentum ($k \neq k'$). Therefore, the non-ideality in the noise
suppression observed in Ref.~\onlinecite{EC} cannot be caused by
the SO coupling.

\subsection{Spin-polarized electron collision}
In the previous example, an unpolarized collision cannot show any
SO coupling effect, because, starting with two quiet sources of
electrons with fully occupied energy levels, all the transport
statistics are governed by the Pauli exclusion principle
independent of the SO coupling. However, this is not the case for
a spin-polarized collision. As the standard spin basis is not the
stationary basis, and as the SO coupling scattering matrix mixes
different SO coupling states, some of the partition noise is
recovered, depending on the SO coupling constant $\alpha$ and on
the angle between the colliding leads. We will confirm this by
calculating the non-equilibrium noise (of the electrons above the
Fermi sea) between $E_{f} $ and $E_{f} + eV$, assuming that the
process of polarization does not effect the transport in the
conductor (for example, there is no magnetic field in the
conductor). The input state is then made of all the electrons with
the same spin, for example spin-up, between $k_{f}$ and
$k_{\max}=k(E_{f}+eV)$. Following Ref.~\onlinecite{LB}, we present
the initial state and the current operator using the following
notations
\begin{eqnarray} \left| \Psi _{i}\right>
& = & \prod_{k=k_{f}}^{k_{max}} \hat{a}^{\dag}_{1 \uparrow}(k)
\hat{a}^{\dag}_{2 \uparrow}(k)\left| 0\right> \\
\hat{I}_{\alpha}(t)& = & \frac{e}{h}\sum_{s}\sum_{\sigma
\sigma'}\sum_{\beta \gamma}\int dE dE'
e^{\frac{i(E-E')t}{\hbar}}A^{\alpha s}_{\beta \gamma \sigma
\sigma'}\nonumber \\* & & \qquad \qquad \qquad \quad
\times\;\hat{a}^{\dag}_{\beta \sigma}(E)\hat{a}_{\gamma
\sigma'}(E') \\ A^{\alpha s}_{\beta \gamma \sigma \sigma'} & = &
\delta_{\alpha \beta}\delta_{\alpha \gamma}\delta_{s
\sigma}\delta_{s \sigma'} - S^{* \alpha s}_{\beta \sigma}S^{\alpha
s} _{\gamma \sigma'}
\end{eqnarray}
where $k_{max}$ is given by $E_{f} + eV = \frac{\hbar^{2}
k_{max}^{2}}{2m}$. We then use the current fluctuation $\delta
\hat{I}_{\alpha}(t) = \hat{I}_{\alpha}(t) -
\left<\hat{I}_{\alpha}(t)\right>$ to calculate the fluctuation
correlation function.
\begin{widetext}
\begin{eqnarray}
\left<\delta \hat{I}_{\alpha}(t)\delta \hat{I}_{\alpha}(0)\right>
& = & \frac{e^{2}}{h^{2}}\sum_{s s'}\sum_{\sigma \sigma' \sigma''
\sigma '''}\sum_{\beta \gamma \delta \zeta} \int dE dE' dE'' dE'''
e^{\frac{i(E-E')t}{\hbar}}A^{\alpha s}_{\beta \gamma \sigma
\sigma'}A^{\alpha s'}_{\delta \zeta \sigma'' \sigma'''}\;\Delta
^{\sigma \sigma' \sigma'' \sigma '''}_{\beta \gamma \delta \zeta}
\label{eq15}  \\
 \Delta ^{\sigma \sigma' \sigma'' \sigma '''}_{\beta
\gamma \delta \zeta} & \equiv & \left<\hat{a}^{\dag}_{\beta
\sigma}(E)\hat{a}_{\gamma \sigma'}(E')\hat{a}^{\dag}_{\delta
\sigma''}(E'')\hat{a}_{\zeta \sigma'''}(E''')\right> -
\left<\hat{a}^{\dag}_{\beta \sigma}(E)\hat{a}_{\gamma
\sigma'}(E')\right>\left<\hat{a}^{\dag}_{\delta
\sigma''}(E'')\hat{a}_{\zeta \sigma'''}(E''')\right>
\end{eqnarray}
\end{widetext} Although we derived an expression for the current
operator in the SO coupling basis, it is mathematically more
convenient in this case to express all the operators in the
standard spin basis
\begin{equation}
\hat{a}^{\dag}_{\beta \sigma}(E)  =  \frac{[\;e^{i
\frac{\theta_{\beta}}{2}} \hat{a}^{\dag}_{\beta \uparrow
}(k(E,\sigma)) + \eta_{\sigma} e^{-i
\frac{\theta_{\beta}}{2}}\hat{a}^{\dag}_{\beta \downarrow
}(k(E,\sigma))\;]}{\sqrt{2 D_{\sigma}(k)}}
\end{equation}
with $D_{\sigma}(k) \equiv \frac{dE(k,\sigma)}{dk} =
\frac{\hbar^{2} k}{m} + \eta_{\sigma} \alpha $. Defining $\Pi
\equiv \sqrt{D_{\sigma}(k)
D_{\sigma'}(k')D_{\sigma''}(k'')D_{\sigma'''}(k''')}$, we have,
\begin{widetext}
\begin{eqnarray}
\Delta ^{\sigma \sigma' \sigma'' \sigma '''}_{\beta \gamma \delta
\zeta}  & = &  \frac{1}{4\Pi} \; \Big[  \; \delta_{\beta \zeta}
\delta_{\gamma \delta} \;\delta (k_{E,\sigma} - k_{E''',\sigma'''}
)\; \delta (k_{E',\sigma'} - k_{E'',\sigma''})\; n_{\beta \uparrow
}(k_{E,\sigma})\;[1-n_{\gamma \uparrow}(k_{E',\sigma'})] \;
\nonumber \\* \!\!\!\!\!\!\!\!\!\!\!\!\! & \!\!\!\!\!\! &
\!\!\!\!\!\! \qquad \qquad  ...+\;\eta_{\sigma'}\eta_{\sigma''}
\delta_{ \beta \zeta } \delta_{\gamma \delta} \;\delta
(k_{E,\sigma} - k_{E''',\sigma'''}) \delta (k_{E',\sigma'} -
k_{E'',\sigma''}) \; n_{\beta \uparrow}(k_{E, \sigma})\; \Big]
\end{eqnarray}
where $k_{E,\sigma} = k_{E',\sigma'}$ for $E = E' + \alpha k (\eta
_{\sigma'} - \eta_{\sigma})$, so that $ \delta( k_{E,\sigma} -
k_{E''',\sigma'''}) = D_{\sigma}(k) \delta[E- E''' + \alpha
k(\eta_{\sigma} - \eta_{\sigma'''})]$. After integration over
$E^{''}$ and $E^{'''}$ and making the approximation $D_{+}(k)
\approx D_{-}(k)$, we find
\begin{eqnarray}
 \int dE'' dE''' \Delta
^{\sigma \sigma' \sigma'' \sigma '''}_{\beta \gamma \delta \zeta}
& = & \frac{1}{4} \Big[ \; \delta_{\beta \zeta} \delta_{\gamma
\delta}  \; n_{\beta \uparrow }(k_{E,\sigma})\;[1-n_{\gamma
\uparrow}(k_{E',\sigma'})] \; + \;\eta_{\sigma'}\eta_{\sigma''}\;
\delta_{\gamma \delta} \; n_{\beta \uparrow}(k_{E, \sigma})\;
\Big] . \label{eq16}
\end{eqnarray} \end{widetext}
\pagebreak
We can replace $D_{-}(k)$ by $D_{+}(k)$, because
$\frac{D_{-}(k)}{D_{+}(k)} = \frac{\frac{\hbar^{2} k}{m} -
\alpha}{\frac{\hbar^{2} k}{m} +\alpha} \approx 1 +
\frac{\alpha}{\frac{\hbar^{2} k}{2 m}} \approx 1$, which follows
from $\frac{\alpha}{\frac{\hbar^{2} k}{2 m}}\leq \frac{\alpha
k_{f}}{\frac{\hbar^{2} k_{f}^{2}}{2m}} =
\frac{E_{SO}(k_{f})}{E_{c}(k_{f})} \ll 1 $ as the SO coupling is
small. We will now calculate the power spectral density at zero
frequency in output lead number 3, $S_{33}(0)$,
\begin{equation}
S_{33 }(\omega) \equiv 2\;\int dt \; e^{i \omega t} \left< \delta
I_{3}(t) \; \delta I_{3}(0)\right> .
\end{equation}
From Eq.~(\ref{eq15}) and Eq.~(\ref{eq16}), we find
\begin{eqnarray}
S_{33 }(0) =  \frac{e^{2}}{2 h}  \sum_{s s'} \sum_{\sigma \sigma'
\sigma'' \sigma'''} \sum_{\beta , \gamma = 1, 2} A^{3 s}_{\beta
\gamma \sigma \sigma'} \;A^{3 s'}_{ \gamma \beta \sigma''
\sigma'''} \nonumber \\
\times \int dE \;n_{\beta \uparrow}(k_{E,\sigma}) [ 1 - n_{\gamma
\uparrow}(k_{E,\sigma'}) + \eta_{\sigma'}\eta_{\sigma''}]
\label{eq17}
\end{eqnarray}
where $\gamma =$ 1 or 2, since we ignored scattering from output
to output. $n_{\beta \uparrow }(k_{E,\sigma})$ is the number of
electrons in lead $\beta$ with spin up and momentum $k_{E,\sigma}$
\[
\begin{array}{l}
n_{\beta \uparrow }(k_{E,\sigma})  = \\ \\
      \;     \left\{ \begin{array}{l}
           0 \;\; \text{ if } \;\; E \leq E_{f} + \eta_{\sigma}
           \alpha k_{f} \\*  1 \;\; \text{if} \;\; E_{f}
           + \eta_{\sigma} \alpha k_{f} \leq E \leq E_{f} + eV
           + \eta_{\sigma} \alpha k_{max} \\*
           0 \;\; \text{if} \;\;E_{f} + eV + \eta_{\sigma}
           \alpha k_{max} \leq E
           \end{array} \right. . \\
\end{array} \]
Assuming that the SO coupling is weaker than the bias voltage,
that is, $E_{f} + \alpha k_{f} \leq E_{f} + eV - \alpha k_{max}$,
we have $n_{\beta \uparrow}(k_{E,-})[1-n_{\gamma
\uparrow}(k_{E,+})] \neq 0$  for $E_{f} - \alpha k_{f} \leq E \leq
E_{f} + \alpha k_{f}$ and $n_{\beta \uparrow
}(k_{E,+})[1-n_{\gamma \uparrow}(k_{E,-})] \neq 0$ for $E_{f} + eV
- \alpha k_{max} \leq E \leq E_{f} + eV - \alpha k_{max}$.
Therefore, Eq.~(\ref{eq17}) becomes
\begin{widetext}
\begin{equation} S_{3 3 }(0) = \frac{ e^{2} }{2 h}
\sum_{s s'} \sum_{\sigma \sigma' \sigma'' \sigma'''} \sum_{\beta ,
\gamma} A^{3 s}_{\beta \gamma \sigma \sigma'} \;A^{3 s'}_{ \gamma
\beta \sigma'' \sigma'''} [\;\eta_{\sigma'} \eta_{\sigma''}\; eV +
\delta_{\sigma +} \delta_{\sigma' -}\;2\; \alpha k_{max} +
\delta_{\sigma -} \delta_{\sigma' +}\;2\; \alpha k_{f} ] .
\end{equation}
\end{widetext}
Given the scattering matrix calculated in Eq.~(\ref{eq14}), we
find
\begin{equation}
\sum_{s s'} \sum_{\sigma \sigma' \sigma'' \sigma'''} \sum_{\beta ,
\gamma } A^{3 s}_{\beta \gamma \sigma \sigma'} \;A^{3 s'}_{ \gamma
\beta \sigma'' \sigma'''} \eta_{\sigma'}\eta_{\sigma''} = 0 ,
\end{equation}
and \begin{equation} \sum_{s s'} \sum_{\sigma \sigma' \sigma''
\sigma'''} \sum_{\beta , \gamma} A^{3 s}_{\beta \gamma \sigma
\sigma'} \;A^{3 s'}_{ \gamma \beta \sigma'' \sigma'''}
\delta_{\sigma +} \delta_{\sigma' -} = 4 \;T R \sin^{2} \theta ,
\end{equation}
so that
\begin{equation}
S_{3 3}(0)  =  \frac{4e^{2}}{h} \;(\alpha k_{f} + \alpha
k_{max})\;T (1-T) \sin^{2} \theta .
\end{equation}
After some calculation, we find the current and Fano factor to be
\begin{eqnarray}
\left< I_{3}\right> & = & \frac{e^{2} V}{h} \\ F & \equiv &
\frac{S_{33}(0)}{\left<I_{3}\right>} \; = \; 4 \;e T(1-T) \sin^{2}
\theta \;\frac{ \alpha (k_{f} + k_{max})}{eV} . \nonumber \\*
\end{eqnarray}
This result reveals two typical features of SO coupling: first,
the noise is proportional to the SO coupling constant, and second,
it depends on the angle between the input leads 1 and 2. One can
check that the obtained noise is identical to the partition noise
for two independent leads where only a fraction
($\frac{\alpha(k_{f} + k_{max})}{eV})$ of the electrons is
colliding, and for which the noise is modified by the factor
$\sin^{2} \theta$.
\begin{figure}[h]
\epsfig{file=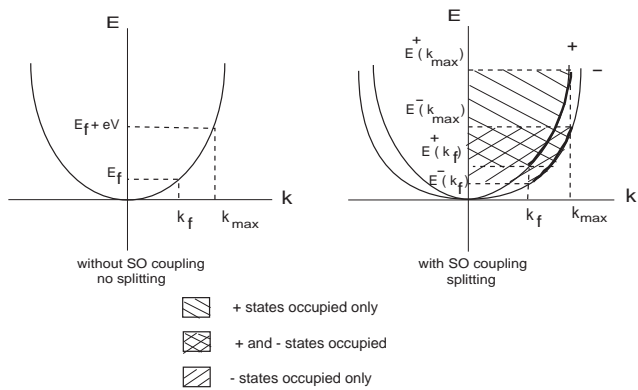,height=2in} \caption{Energy diagrams with
and without SO coupling} \label{Fig6}\end{figure} This result can
also be explained using Fig.~\ref{Fig6}. In the inputs, the
spin-up electron states between $k_{f}$ and $k_{max}$ are fully
occupied. However, in terms of the SO coupling states, only the
states between $E(k_{f}, +)$ and $E(k_{max},-)$ are jointly
occupied, giving no contribution to the noise as in the
unpolarized case. Between $E(k_{f},-)$ and $E(k_{f},+)$, only the
$-$ states are filled, and between $E(k_{max},-)$ and
$E(k_{max},+)$, only the $+$ states are filled. These states do
not have the same spin in leads 1 and 2 (their overlap is
$\cos^{2}\theta$), so that the Pauli exclusion principle does not
fully suppress the noise, and the classical partition noise is
partially recovered through the factor $\sin^{2}\theta =
1-\cos^{2}\theta$. In this treatment, we have assumed that the SO
coupling is small ($ \frac{\alpha (k_{f} + k_{max})}{eV} \leq 1$),
so that $E(k_{f}, +) \leq E(k_{max},-)$. But, a typical value of
the applied bias voltage is $eV = 0.1$ meV, which is smaller than
the typical values of energy splitting caused by the Rashba effect
($0.3$ meV $\leq 2 \alpha k_{f} \leq 3$ meV). In this case, all
the electrons contribute to the partition noise, and we have $F =
4 \; eT(1-T) \sin^{2} \theta$. Depending on the angle between the
leads, we can go from full noise suppression to the classical
limit of the partition noise. The modification of the noise caused
by SO coupling in this collision experiment can in principle be
measured, even for small values of the SO coupling constant.

\section{Conclusion}
We have studied the influence of SO coupling in the framework of
the Landauer-B\"{u}ttiker formalism. A short review of the effect
of SO coupling in 2DEGs (and more precisely of the Rashba effect)
has reminded us of its main features: spin-splitting proportional
to $k$, and stationary states of the spin perpendicular to the
direction of propagation. The electron feels a magnetic field
perpendicular to the direction of propagation, with an amplitude
proportional to the velocity. We have then included the effects of
SO coupling in the Landauer-B\"{u}ttiker coherent scattering
formalism. The SO coupling gave rise to two important
modifications. First, the addition of a SO coupling term in the
Hamiltonian modifies the expression of the current operator, and
an extra term directly related to SO coupling has to be included.
Second, the expansion of the current operator is in the basis of
the stationary states of SO coupling. The final formula for the
current operator was found to be identical to the one derived in
the spin-independent transport case, but with the spin replaced by
the SO coupling label, indicating the alignment of the spin on the
virtual magnetic field caused by the SO coupling. The main
differences introduced by the SO coupling then arise in the
calculation of the scattering matrix relating the stationary
states of SO coupling in different leads with different
orientations. The direction of the virtual magnetic field (and,
consequently, the direction of the spin) depends on the direction
of propagation which is different for each lead in general.
Therefore, the scattering matrix is shown to mix states with
different SO coupling labels, and the strength of this mixing
depends on the angle between the leads. The effect of SO coupling
on the current noise was then investigated in two examples of
electron collision. In the unpolarized electron collision example,
it is shown that the SO coupling does not modify the noise; this
case is entirely determined by the Pauli exclusion principle. In
contrast, the polarized case exhibits a contribution to the noise
caused by SO coupling, which is proportional to the SO coupling
constant $\alpha$ and depends on the angle between the leads. A
polarized electron collision experiment provides, in principle,
another way to measure the strength of the Rashba splitting
energy. Finally, this new formulation of the current operator can
be applied to other coherent scattering experiments in which one
wants to investigate or incorporate the effects of SO coupling.

\begin{acknowledgments}
The authors gratefully acknowledge useful discussions with E.
Waks, X. Ma\^{i}tre and the support of J. F. Roch.  We thank the
Quantum Entanglement Project (ICORP, JST) for financial support.
W. D. O. gratefully acknowledges additional support from MURI and
the NDSEG Fellowship Program.
\end{acknowledgments}

\appendix
\section{Density of states with spin-orbit coupling} \label{apa}
 In this appendix, we calculate the density of states
and show that it does not depend on the SO coupling label
$\sigma$, as is suggested by the symmetry between the $+$ and $-$
states in the energy dispersion diagram (see Fig.~\ref{Fig2}),
\begin{eqnarray}
E &=& \frac{\hbar^{2} k^{2}(E,\sigma)}{2 m} + \eta_{\sigma} \alpha
k(E,\sigma) \\ v_{g \sigma}(E) & = & \frac{1}{\hbar}\frac{d
E(k,\sigma)}{d k} \; = \; \frac{\hbar \; k(E,\sigma)}{m}
+\eta_{\sigma} \frac{\alpha}{\hbar} .
\end{eqnarray}
From Eq.~(\ref{eq4}) we deduce
\begin{eqnarray}
k(E,\sigma) & = & k(E,+) + (\eta_{\sigma}-1) \frac{m \alpha }{
\hbar^{2}} \\ v_{g \sigma}(E) & =& \frac{\hbar k(E,+)}{m} +
\frac{\alpha}{\hbar} \;=\; v_{g}(E)\\ \rho(E) & =
&\frac{\rho(k)}{\hbar \; v_{g}(E)} \;=\; \frac{L}{2 \pi}
\frac{1}{\hbar \;v_{g}(E)}
\end{eqnarray}
which is independent of $\sigma$.
\section{Scattering matrix in the four-port beamsplitter case} \label{apb}
\begin{figure}[h]
\epsfig{file=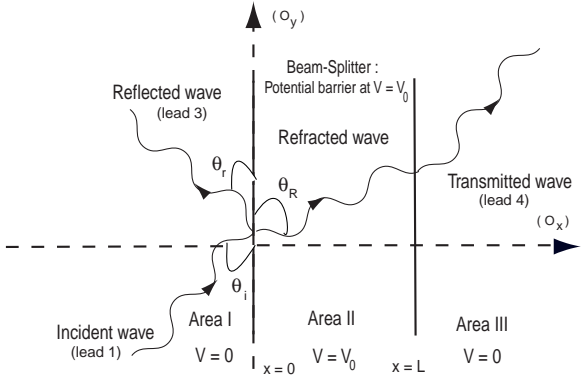,height=2in} \caption{Reflected, refracted
and transmitted waves in a beamsplitter} \label{Fig7}
\end{figure}

\begin{figure}[h]
\epsfig{file=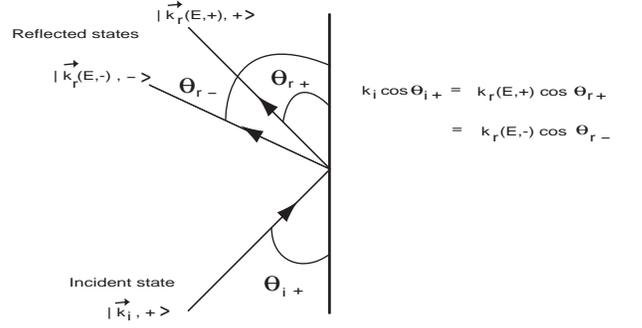,height=1.7in} \caption{angular separation
after reflection at a beamsplitter} \label{Fig8}
\end{figure}

Here, we determine the scattering matrix for the four-port
beamsplitter described in Fig.~\ref{Fig5}. The beamsplitter is
very simply approximated by a potential barrier at $V=V_{0}$ of
length $L$. The plane is then divided into three areas of
different potential (see Fig.~\ref{Fig7}) in which the solution of
the Schr\"{o}dinger equation is known.
Starting with an incident wave in lead number one, for example, we
can calculate the reflected, refracted, and transmitted waves in
lead 3 and 4, using the continuity of the wavefunction and its
derivatives at the beamsplitter interface (x=0 and x=$L$). For
example, let us start with an incident state at energy $E$ with
momentum $k_{i}$ in the SO coupling state $\sigma=+$. By the
conservation of energy, the reflected wave in lead 3 is a
superposition of the states $ \left| \mathbf{k_{r}}(E,+),
+\right>$ and $\left| \mathbf{k_{r}}(E,-), -\right>$ with
\begin{equation}
k_{r}(E,+) = k_{i}(E,+) = k_{r}(E,-) - \Delta k .
\end{equation}
The translational invariance of the beamsplitter along the y-axis
leads to the conservation of the y-component of the momentum
\begin{equation}
k_{i} \cos \theta_{i} = k_{r}(E,+) \cos \theta_{+} = k_{r}(E,-)
\cos \theta_{-} .
\end{equation}
We deduce that $\cos \theta_{i} = \cos \theta_{r +}$, but $\cos
\theta_{i} \neq \cos \theta_{r -}$. There is dispersion due to the
SO coupling, leading to an angular separation between the $+$ and
$-$ states after reflection at the beamsplitter (see
Fig.~\ref{Fig8}).
This angular separation is given by
\begin{equation}
\theta_{r -} = \arccos[ (1 - \frac{\Delta k}{k_{r}(E,-)})
\cos\theta_{i}] \neq \theta_{r +} .
\end{equation}
Starting with an incident state with SO coupling label $-$, the
angular separation is
\begin{equation}
\theta_{r +} = \arccos[ (1 + \frac{\Delta k}{k_{r}(E,+)})
\cos\theta_{i}] \neq \theta_{r -} .
\end{equation}
In analogy with the the total reflection for incident angles below
the critical angle in classical optics, we can even have $(1 +
\frac{\Delta k}{k_{r}(E,+)}) \cos\theta_{i} >1$, leading to a
suppression of the reflection in the $+$ state for a small enough
incident angle. We note that starting with a mixture of $+$ and
$-$ states in the incident beam of electrons, one could suppress
the reflection of the $+$ state, thus achieving a polarization of
the beam. Although interesting, we will neglect this effect of
angular dispersion by considering only non-equilibrium electrons
above the Fermi energy for which $\frac{\Delta k}{k} \ll 1$ (as SO
coupling is small compared to the kinetic energy), and important
incident angles $(\theta_{i} = \frac{\pi}{4})$. In this case the
angular separation is very small ($\Delta \theta = \theta_{r -} -
\theta_{r +} \ll 1$). The equations of continuity of the
wavefunction and its derivative are then much easier to solve, and
one can find that the incident, refracted, and reflected waves
have the same spin at x=0, and the refracted and transmitted waves
have the same spin at x=$L$. Using Eq.(\ref{eq2}) to find the spin
overlap between different leads we find, starting with a $+$
incident state,
\begin{equation}
\left| \Psi_{t}\right> = t \left| k,+\right> \;\;\;\; \left|
\Psi_{r}\right> = r \;[ \cos \theta_{i} \left| k, + \right> + \;i
\sin\theta_{i} \left| k,-\right>]
\end{equation}
where $\left| \Psi_{t}\right>$ is the transmitted wave (into lead
4) and $\left| \Psi_{r}\right>$ the reflected wave (into lead 3).
As the transmitted wave has the same direction of propagation as
the incident wave, there is no mixing of the SO coupling states,
and we have only the usual transmission coefficient $t$. For the
reflected wave, the direction is changed and we have to mix the
different SO coupling states to obtain the same spin as the
incident wave on the interface with the beamsplitter. If we start
now with a $-$ incident state, we find
\begin{equation}
\left| \Psi_{t}\right> = t \left| k,-\right> \;\;\;\;\left|
\Psi_{r}\right> = r \;[ i \sin\theta_{i} \left| k, + \right> + \;
\cos \theta_{i} \left| k,-\right>] .
\end{equation}
The same analysis can be done for an incident state in lead 2 with
$\theta_{i}$ replaced by $-\theta_{i}$. We then deduce the whole
scattering matrix:
\begin{equation}
S = \left[ \begin{array}{cccc} r \cos\theta_{i} & i
r\sin\theta_{i} & t & 0 \\ i r\sin\theta_{i} & r \cos\theta_{i} &
0 & t \\ t & 0 & r \cos\theta_{i} & -i r\sin\theta_{i} \\ 0 & t &
-i r\sin\theta_{i} & r \cos\theta_{i}
\end{array} \right]
\end{equation}

\end{document}